\documentclass{sn-jnl}

\usepackage{graphicx}%
\usepackage{multirow}%
\usepackage{amsmath,amssymb,amsfonts}%
\usepackage{amsthm}%
\usepackage{mathrsfs}%
\usepackage[title]{appendix}%
\usepackage{xcolor}%
\usepackage{textcomp}%
\usepackage{manyfoot}%
\usepackage{booktabs}%
\usepackage{algorithm}%
\usepackage{algorithmicx}%
\usepackage{algpseudocode}%
\usepackage{listings}%
\usepackage{siunitx}
\usepackage{nicefrac}
\usepackage{anysize} 
\usepackage{gensymb}

\marginsize{2.5cm}{2.5cm}{2cm}{2cm}

\begin{document}

\title{Confinement in metal-organic frameworks as a route to harnessing liquid barocalorics in the solid-state}

\author[1]{\fnm{Ming} \sur{Zeng}}

\author[2]{\fnm{Frederic} \sur{Rendell-Bhatti}}

\author[3]{\fnm{Eamonn} \sur{T. Connolly}}%

\author[4]{\fnm{Yang} \sur{Wang}}%

\author[1]{\fnm{Josep-Lluís} \sur{Tamarit}}%

\author[4]{\fnm{Ross} \sur{S. Forgan}}%

\author*[1]{\fnm{Pol} \sur{Lloveras}}\email{pol.lloveras@upc.edu}

\author*[2]{\fnm{David} \sur{Boldrin}}\email{david.boldrin@glasgow.ac.uk}%

\affil[1]{\orgdiv{Group of Characterization of Materials, Department of Physics and Barcelona Research Center in Multiscale science and Engineering}, \orgname{Universitat Politècnica de Catalunya}, \orgaddress{\street{Av. Eduard Maristany 10-14}, \postcode{08019}, \city{Barcelona}, \state{Catalonia}, \country{Spain}}}

\affil[2]{\orgdiv{SUPA, School of Physics and Astronomy}, \orgname{University of Glasgow}, \orgaddress{\city{Glasgow} \postcode{G12 8QQ}, \country{United Kingdom}}}

\affil[3]{\orgdiv{Diamond Light Source, Diamond House}, \orgname{Harwell Science and Innovation Campus}, \orgaddress{\street{}, \postcode{OX11 0DE} \city{Didcot, Oxforshire}, \country{United Kingdom}}}

\affil[4]{\orgdiv{School of Chemistry}, \orgname{University of Glasgow}, \orgaddress{ \city{Glasgow} \postcode{G12 8QQ}, \country{United Kingdom}}}

\maketitle

\begin{abstract}

Barocaloric (BC) effects at liquid-vapor transitions in hydrofluorocarbons drive most commercial technologies used for heating and cooling in the heating, ventilation and air-conditioning sector. However, these fluids suffer from huge global warming potential and alternative gases are less efficient, toxic or flammable. Solid-solid and solid-liquid BC materials have zero global warming potential and could even improve on current device efficiencies. Whilst solid-liquid BCs typically outperform solid-solid BCs, the latter are advantageous as they avoid leaks and present easier handling and recyclability thus facilitating waste management. Here we confine the solid-liquid BC stearic acid inside the nanopores of a functionalized metal-organic framework (MOF) and demonstrate that the colossal BC properties are retained in a solid-state material. Moreover, the enhanced interactions between the pore surface and the BC material allow a level of active control over the thermal response, as opposed to passive encapsulation. Our results open novel avenues to exploit and tune colossal BC effects in a wide range of combinations of solid-liquid BC materials embedded within functionalized MOFs, without the associated engineering drawbacks.

\end{abstract}

\section{Introduction}

Conventional refrigerants used in vapour compression technologies for cooling and heating are commonly based on hydrofluorocarbons (HFCs), which display very high global warming potential (GWP) when released to the atmosphere. Current alternatives, such as natural refrigerants, hydrocarbons or hydrofluoroolefins, suffer from other hazards such as toxicity, flammability or reduced efficiency \cite{EuropeanComission}. Since they are all gases, leakage, containment and recycling at a product’s end-of-life is challenging. These drawbacks are further exacerbated by the widespread use of these technologies, so even small improvements could therefore have a significant impact. Solid-state refrigerants are low-GWP, safer and easily contained, and thus, are widely regarded as promising next-generation alternatives for cooling and heating \cite{moya2020caloric}.

\medskip
Thermal effects in solids can be driven by different external stimuli such as magnetic, electric and/or mechanical fields, and may become large near first-order phase transitions (FOPTs). Barocalorics (BCs) \cite{lloveras2023barocaloric}, which are operated using hydrostatic pressure, have so far demonstrated the largest isothermal entropy changes among solids, approaching those of conventional HFCs. These colossal BC effects have been observed in a wide range of materials\cite{lloveras2021advances,boldrin2021fantastic} such as plastic crystals, hybrid organic-inorganic compounds and spin-crossover complexes, some of which were previously proposed as phase change materials (PCMs) for thermal energy storage \cite{ahvcin2024latent}. The great capacity to store and exchange heat in these compounds emerges due to a FOPT from a crystalline ordered phase towards another crystalline phase, yet with significant molecular dynamic disorder. Notably, the easy manipulation of the transition via hydrostatic pressure relies on a large volume difference between the two phases. 

\medskip
Following further inspiration in PCMs, recent studies have reported comparable or even larger BC effects at solid-liquid (SL) transitions in a variety of compounds such as \textit{n}-alkanes \cite{lin2022colossal,miliante2022colossal}, poly(ethylene glycol) \cite{yu2022colossal}, copolymers \cite{zhang2022giant} and fatty acids \cite{xiong2024colossal,imamura2025colossal}. Their low cost, high availability, good cyclability and relative non-toxicity increase their technological appeal. However, the liquid state negates some of the benefits provided by solids. For this reason, the mechanical encapsulation of SL PCMs is an attractive solution to retain the benefits of SL barocaloric transitions within a solid material. Encapsulation of SL PCMs has recently been demonstrated within a passive microporous matrix \cite{Shuang2025Colossal}. Here, we show encapsulation of SL BCs in the nanoporous structure of a metal-organic framework (MOF), enabling both the passive containment of a liquid phase in addition to active control of its BC properties. As functional groups in modified MOFs have been found to enhance the interplay with adsorbates \cite{lin2012direct}, our case study is realized by incorporating a fatty acid (stearic acid, SA for short) into the amine-functionalized MOF, MIL-101(Cr)-NH$_2$. This strategy, already investigated for thermal energy storage, offers a larger SA loading capacity as compared to canonical MIL-101(Cr) \cite{luan2016introduction,feng2019phase}, despite the decrease in surface area and pore volume that comes with amine functionalisation. This improvement is associated with the formation of hydrogen bonds between amino groups and carboxylates.

\medskip
In the present study, we show that application of hydrostatic pressure of up to 250 MPa does not degrade the confinement of the SL FOPT of SA, thus allowing the SA@MIL-101(Cr)-NH$_2$ composite (SA@MIL for short) to exhibit colossal BC effects in the solid state. Moreover, confinement effects depress the FOPT temperature with respect to pure SA, allowing a level of control over the temperature range of the BC effects in these materials. In addition to exploiting colossal SL BC effects within solid-state devices, our results open new avenues for functionalised control of BC effects, SL or solid-solid, through active interaction within MOFs.

\begin{figure}[h!]
\centering
  \includegraphics[width=0.5\textwidth]{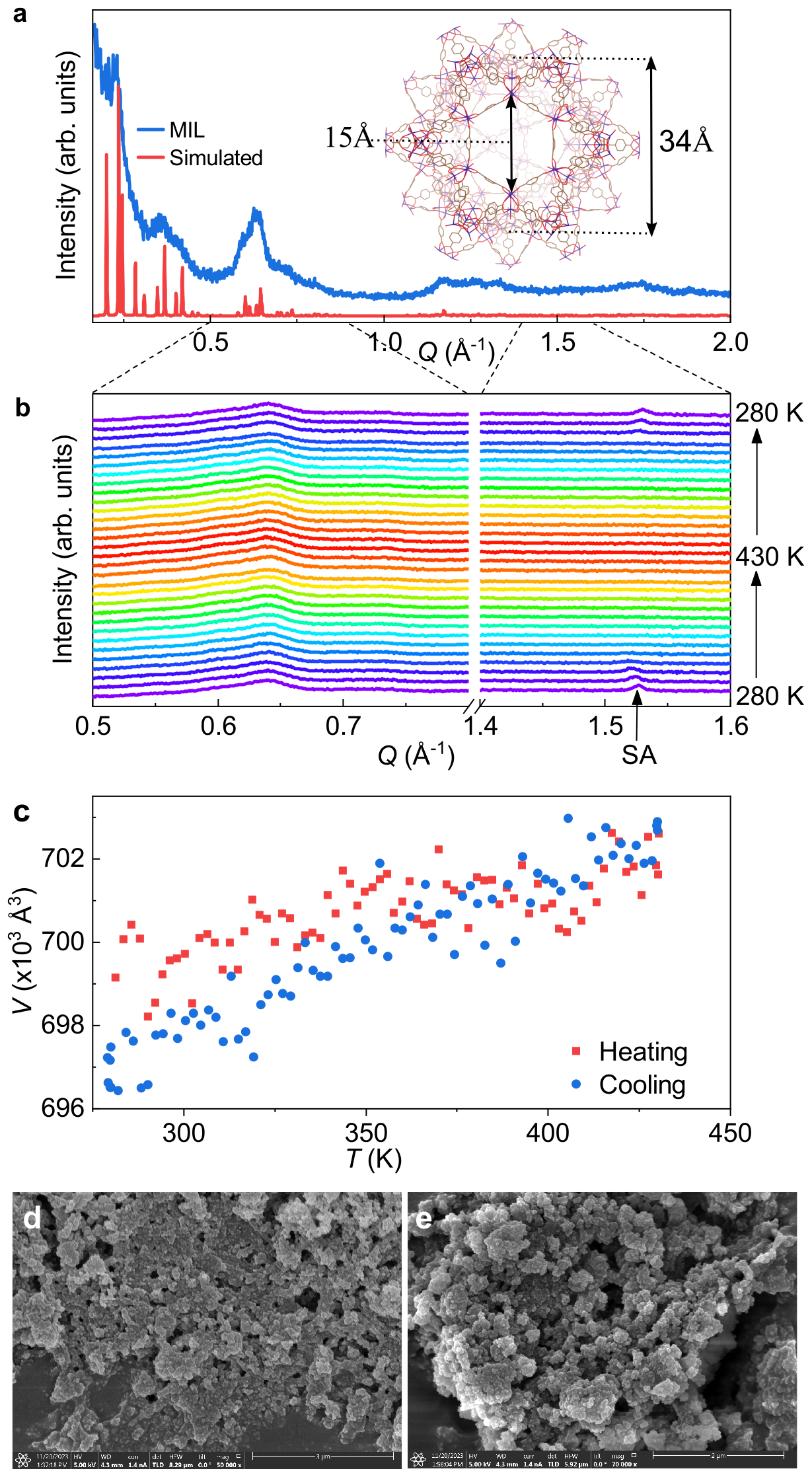}
  \caption{(a) Synchrotron diffraction data on MIL along with a simulated diffraction pattern based on the expected MIL crystal structure. The MIL crystal structure is shown in the inset, with the largest pore sizes indicated; (b) Temperature dependent synchrotron diffraction data; (c) Refined volume from data in (b) of the MIL phase; (d-e) SEM images of (d) $x=0$ and (e) $x=40\%$ of SA@MIL.}
  \label{Fig1}
\end{figure}

\begin{figure}[b!]
\centering
  \includegraphics[width=0.5\textwidth]{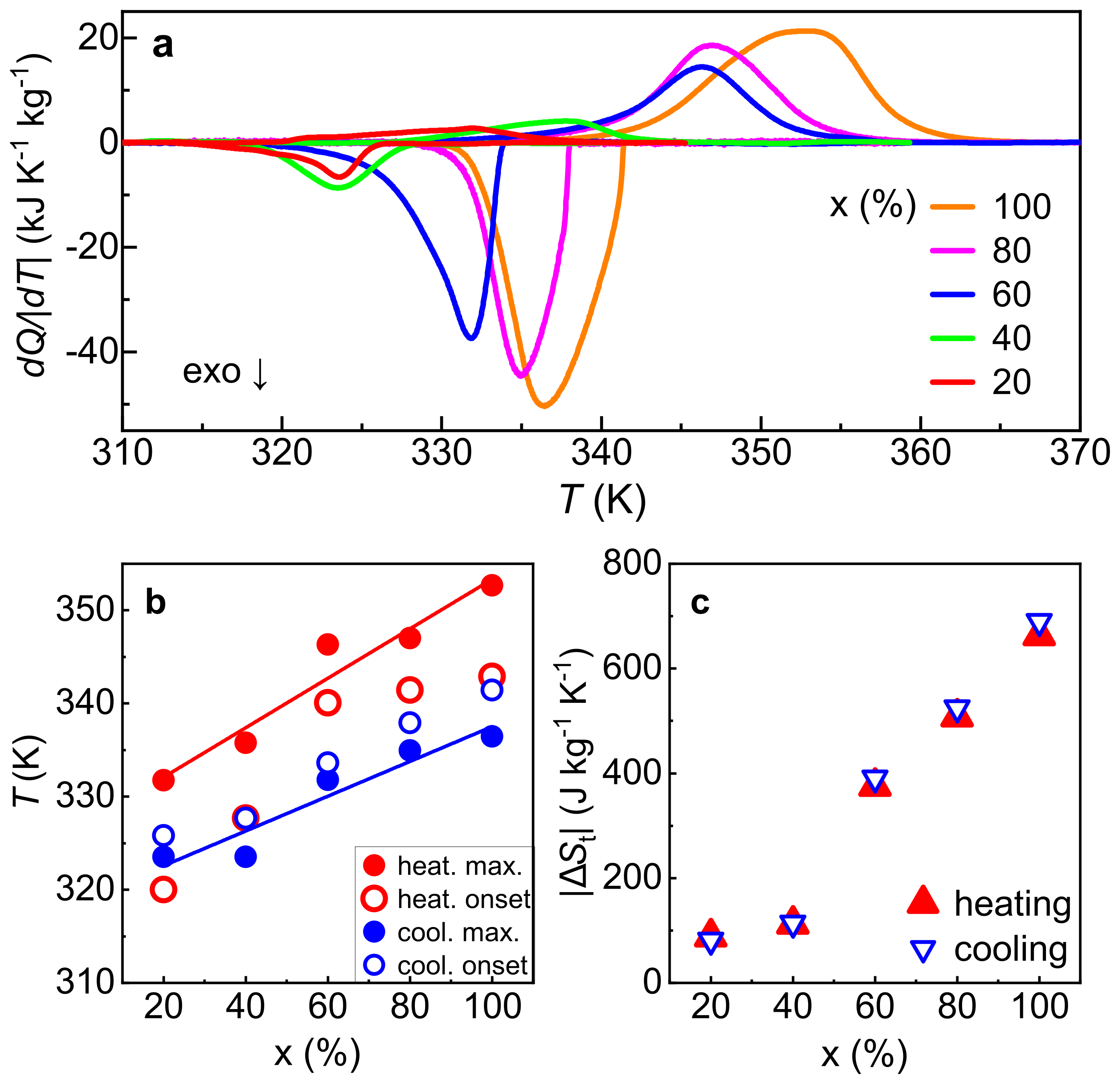}
  \caption{(a) Ambient pressure calorimetric signal as a function of temperature for different SA concentration. (b) Transition temperature as a function of the SA concentration. Solid and empty symbols correspond to peak maxima and onsets, respectively. Lines are fits to peak maxima. (c) Transition entropy change per unit of mass of SA@MIL, as a function of SA concentration.}
  \label{Fig2}
\end{figure}

\section{Results and Discussion}

SA, or octadecanoic acid, with chemical formula CH$_3$(CH$_2$)$_{16}$COOH, is a fatty acid naturally present in living bodies and widely used in the chemical and pharmaceutical industry. As other long chain fatty acids, SA exhibits a number of polymorphs whose overall structure consists of bilayers linked at the carboxyl ends via hydrogen bonds, where the methyl groups facing each other are stabilized via van der Waals interactions \cite{moreno2009competing}. The conformational configuration along the alkyl chain is all \textit{trans} in most cases, with a few polymorphs showing a \textit{gauche} conformation. Structurally, the polymorphs differ from each other mainly by the tilting and direction angle of the aliphatic chain with respect to the crystallographic axes, and by some interatomic distances. At room temperature SA adopts the B form (with a \textit{gauche} conformation near the carboxyl group), which transforms to the C form (all \textit{trans}) at 308-319~K, both forms arranging in the monoclinic space group \textit{P}2$_1$/c with $Z=4$ \cite{moreno2009competing,singleton1950physical,malta1971crystal}. The C form melts at 342.7~K, with a low hysteresis on crystallization, a large latent heat $\Delta H_\text{t}\simeq210$~J~g$^{-1}$ and a large volume change of $\sim$20\%.

\medskip
MIL-101(Cr) arranges in a cubic lattice (space group Fd$\bar{3}$m, $a\sim89$~\AA), exhibiting two types of particularly large cages \cite{ferey2005chromium}. \textbf{Figure \ref{Fig1}a} inset shows the largest of these cages, with a 15\AA\ opening into a cage of 34\AA\ diameter. 
This nanoporous structure of MIL-101(Cr), and MOFs in general, underpins a number of different application opportunities such as CO$_{2}$ capture, H$_{2}$ storage, catalytic activity, and confinement of SL PCMs for thermal energy storage, as has been documented in a number of reviews  \cite{bhattacharjee2014chromium,Zorainy2021,Zou2018}. The amine-functionalized MIL-101(Cr)-NH$_2$ (hereafter referred to as MIL) has the same symmetry and pore structure as MIL-101(Cr).

\medskip
Synchrotron powder X-ray diffraction data of the as-synthesised MIL is shown in Figure \ref{Fig1}a. Compared to the simulated data, the experimental Bragg peaks are significantly broadened due to crystallite size effects, as expected based on the synthetic method \cite{Han2020}. SEM images confirm this, with particle sizes of the order of $\ll$1 $\mu$m (Figure \ref{Fig1}d; see Supplementary Fig. S1 for all other samples). This observation reproduces the published results using the same synthetic method \cite{Han2020} and other literature examples \cite{lin2012direct,Abdpour2020}. Fitting of the data using a Le Bail extraction gives lattice parameters of 88.7 Å, in excellent agreement with the published crystal structure \cite{ferey2005chromium}. The N$_2$ adsorption isotherm (77 K, see Supplementary Figure S2) gave a BET area of 2141~m$^2$~g$^{-1}$ (see Supplementary Figure S3 and S4), in good agreement with expected values \cite{li2017synthesis}.
Synchrotron powder X-ray diffraction data of the as-synthesised MIL is shown in Figure \ref{Fig1}a. Compared to the simulated data, the experimental Bragg peaks are significantly broadened due to crystallite size effects, as expected based on the synthetic method \cite{Han2020}. SEM images confirm this, with particle sizes of the order of $\ll$1 $\mu$m (Figure \ref{Fig1}d; see Supplementary Fig. S1 for all other samples). This observation reproduces the published results using the same synthetic method \cite{Han2020} and other literature examples \cite{lin2012direct,Abdpour2020}. Fitting of the data using a Le Bail extraction gives lattice parameters of 88.7 Å, in excellent agreement with the published crystal structure \cite{ferey2005chromium}. The N$_2$ adsorption isotherm (77 K, see Supplementary Figure S2) gave a BET area of 2141~m$^2$~g$^{-1}$ (see Supplementary Figure S3 and S4), in good agreement with expected values \cite{li2017synthesis}.

\medskip
Upon the encapsulation of SA within the porous structure of MIL, weak unindexed XRD peaks are observed  at $Q = 1.53$ and $1.70$ \AA$^{-1}$, the former indicated by the arrow in the synchrotron powder XRD data in Figure \ref{Fig1}b for $x=40$\%. These peaks align with the 110 and 020 reflections of the expected SA structure, respectively ($P2_{1}/a$, $a=5.587$\AA, $b=7.386$\AA, $c=49.33$\AA, $\beta = 117.24^{\circ}$, CSD 1263280 \cite{Goto2006}). We do not observe any other expected reflections with similar intensity in this $Q$ region and we note that all of these have $l > 0$. The $c$-direction in SA lies along the hydrocarbon chain and therefore the diffraction data suggests that there is no long-range ordering along this direction. We therefore conclude that the diffraction peaks we observe originate from heavily disordered SA within the MIL pores, likely due to an inability for chains to align end-to-end. SEM images of the $x=40$\% loaded sample are shown in Figure \ref{Fig1}e, which looks similar to the pure MIL and provides further evidence that the vast majority of the SA is encapsulated within the MIL.

\medskip
Upon temperature cycling, also shown in Figure \ref{Fig1}b, there are two important features. Firstly, the SA peak disappears at the melting transition ($\sim$325 K) as expected, in agreement with DSC results.  Secondly, the peaks of the MIL phase hardly vary, neither in position nor shape, demonstrating the absence of a phase transition. Indeed, the temperature dependence of the volume of MIL, shown in Figure \ref{Fig1}c, further demonstrates this \cite{lin2012direct,ferey2005chromium}. The latter indicates that the MIL phase should not display a significant BC response. 

\medskip
Ambient pressure calorimetric measurements of the SA@MIL were performed for varying amounts of SA using two setups: DSC and HP-DTA (see Experimental Section). Both setups
display positive and negative peaks in $dQ/|dT|$ associated with the endothermic and exothermic SL FOPT, respectively (see \textbf{Figure \ref{Fig2}a}). Transition temperatures are parameterised from the calorimetric signals by both peak onsets ($T_\text{on}$) and maxima ($T_\text{t}$), to obtain insight on the peak width. Figure \ref{Fig2}b shows $T_\text{on}$ (open symbols) and $T_\text{t}$ (solid symbols) as functions of the SA concentration $x$, as derived from HP-DTA experiments. Interestingly, the SA transition drops by $\sim$20 K from pure SA ($x=100\%$) to $x=20\%$ loading of SA@MIL. Temperature dependence of the SA transition in MIL has previously been observed \cite{luan2016introduction,feng2019phase} and this is likely related to confinement effects, as has been found for a number of aqueous solutions in microporous media \cite{fraux2017forced}. Furthermore, the width in temperature of the calorimetric peaks for all samples where $x < 100$\% are similar to pure SA, suggesting that the calorimetric signal is originating from encapsulated SA@MIL rather than from pure SA. Taken together, the calorimetry, synchrotron diffraction and SEM data all demonstrate that the SA is largely encapsulated within the MIL. 

\medskip
It is worth mentioning that the drop in melting temperature on decreasing the SA load is significantly larger obtained using HP-DTA than that obtained using DSC (see Supplementary Figure S5), the latter in agreement with literature. The discrepancy could occur due to much larger sample masses (which provide more reliable values in the SA concentration) and differences in the sample environment. We based our analysis on the HP-DTA data because the obtained transition temperatures were fully consistent with further measurements using the same setup at high pressures. On the other hand, $T_\text{on}$ displays very small hysteresis as indicated by the difference between red and blue empty symbols suggesting that reversible BC effects may be achievable from very low pressure changes. However, calorimetric peaks are significantly broad even for small scanning rates, so that thermal hysteresis determined from $T_\text{t}$ amounts to $\sim$15~K. This indicates that to harvest reversible BC effects from the full transition, larger pressure changes will be required.

\begin{figure}[t!]
\centering
  \includegraphics[width=0.48\textwidth]{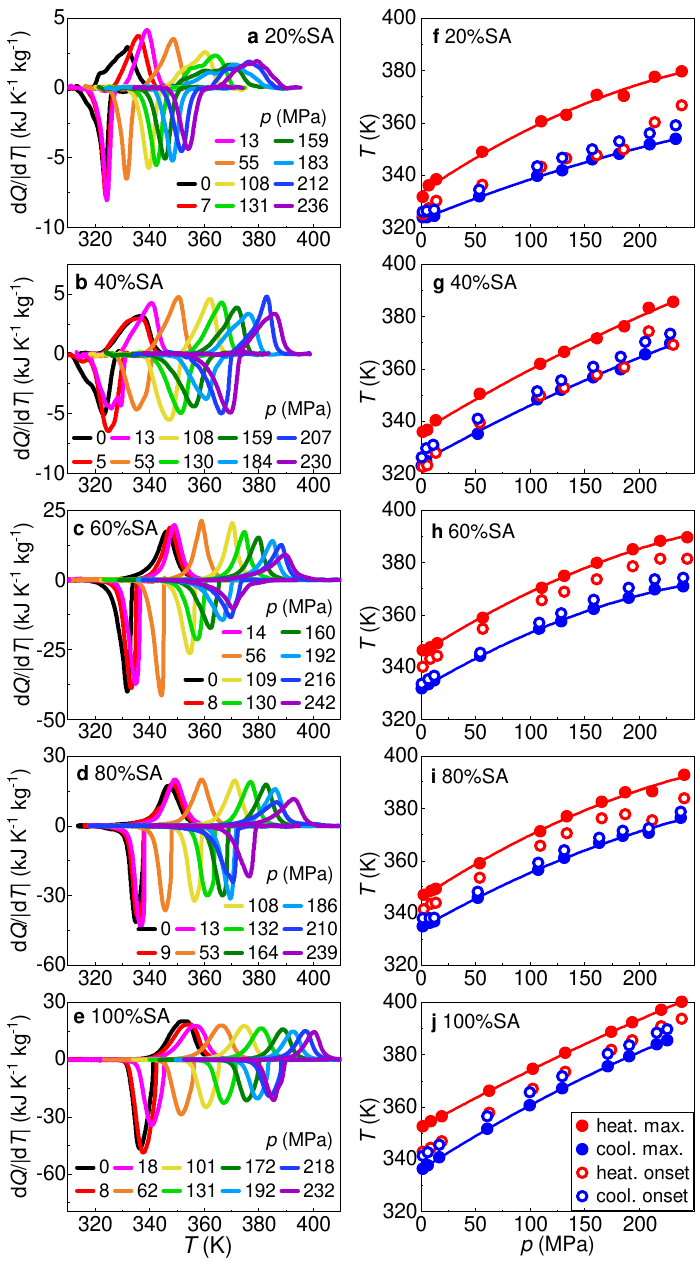}
  \caption{(a-e) High-pressure calorimetric signals measured at different applied pressures and (f-j) as derived respective temperature-pressure phase diagrams in SA@MIL samples with different SA concentration: (a,f) $x=20\%$, (b,g) $x=40\%$, (c,h) $x=60\%$, (d,i) $x=80\%$, (e,j) $x=100\%$. Solid and empty symbols correspond to peak maxima and onsets, respectively. Lines are fits to peak maxima.}
  \label{Fig3}
\end{figure}

\medskip
Changes in enthalpy $\Delta H_\text{t}$ and entropy $\Delta S_\text{t}$ at the FOPT were determined from temperature integrations of $dQ/|dT|$ and $(1/T)(dQ/|dT|)$, respectively, after baseline subtraction, for different SA amounts (see Figure \ref{Fig2}c and Supplementary Figure S6). Pure stearic acid ($x=100$\%) shows $\Delta S_\text{t}\simeq650-700$~J~K$^{-1}$~kg$^{-1}$, in agreement with literature \cite{hasan1994some}, and decreases with decreasing SA concentration, as expected. However, $\Delta S_\text{t}$ is not proportional to the SA amount. For instance, for $x= 20$ and $80$\%, $\Delta S_\text{t}$ is $\sim 15$\% and $\sim75$\% of pure SA, respectively. As has previously been found, this indicates a fraction of crystallization which has been argued to occur due to some proportion of SA molecules in close proximity to and interacting strongly with the MIL pore surface, and therefore not contributing to the entropy change at the SL transition \cite{luan2016introduction,feng2019phase}. The ratio of crystallisation as a function of $x$ for our samples is shown in Supplementary Figure S7. As has been found before, we see a general decrease in this ratio as $x$ is decreased, with $x = 40$\% showing the lowest ratio implying that this loading has the highest proportion of SA interacting strongly with the pore surface. Interestingly, at $x = 20$\% we see this ratio increase again, suggesting that at very low loadings this principle does not hold. However, such low loadings have not been explored before and therefore, given that this is a single data point, further investigations are required to confirm this trend.

\begin{figure}[t!]
\centering
  \includegraphics[width=0.48\textwidth]{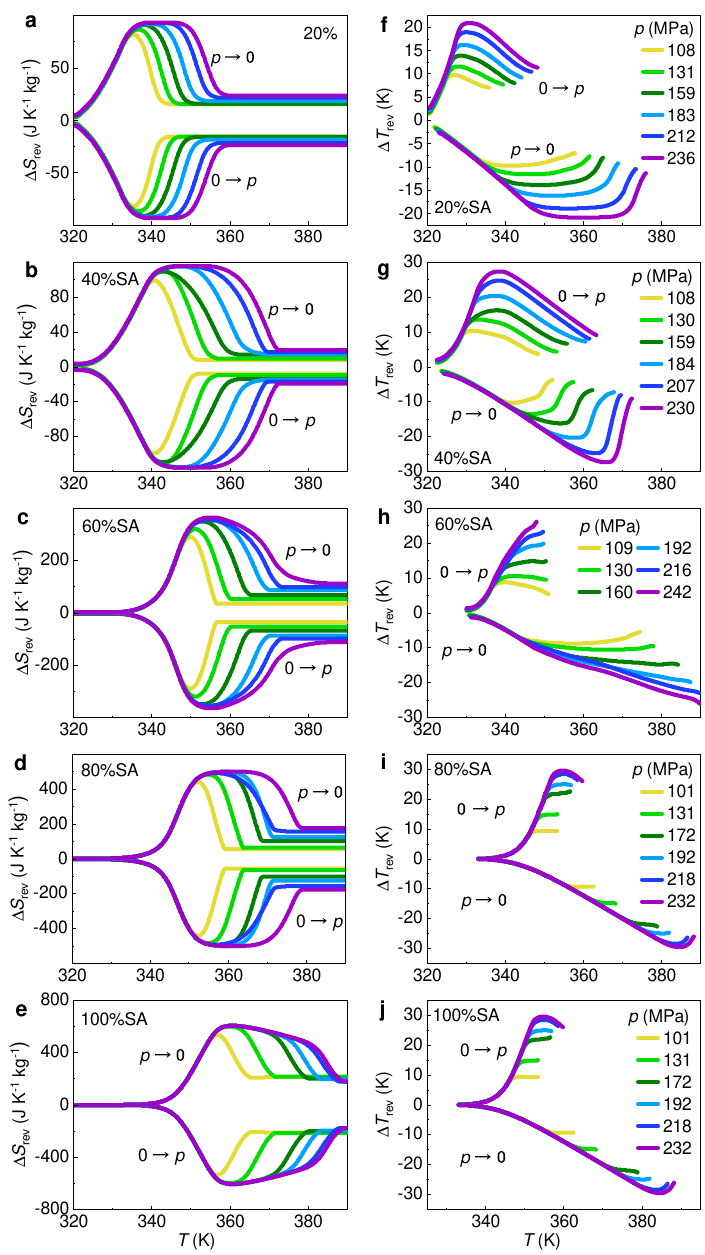}
  \caption{(a-e) Reversible isothermal entropy changes and (f-j) and reverstible adiabatic temperature changes upon pressure changes for different SA concentrations. Color legends in panels (f,g,h,i,j) also hold for panels (a,b,c,d,e), respectively.}
  \label{Fig4}
\end{figure}

\medskip
High-pressure calorimetry measurements were performed on SA@MIL for the same loadings $x=20$, 40, 60, 80 and 100\%. Heat flow peaks after baseline subtraction arising from SA melting for each sample are shown in \textbf{Figure \ref{Fig3}a-e}, from which the temperature-pressure phase diagrams are derived (Figures \ref{Fig3}f-j). Each sample shows a strong dependence of the SA melting temperature on pressure, with $dT/dp$ values ranging between 285 and 230~K~GPa$^{-1}$ for the endothermic transitions at atmospheric pressure (see Supplementary Table S1 for a summary of transition properties). These values decrease at higher pressures due to slight convexity of the coexistence lines. For pure SA, the volume change calculated using the Clausius-Clapeyron equation is $\Delta V_t=\frac{dT}{dp}\Delta S_t\simeq0.15$~cm$^3$~g$^{-1}$, which is in excellent agreement with the reported experimental value \cite{singleton1950physical}. Hysteresis is maintained roughly independent of pressure except for $x=20$\% due to a peak widening. Peak integrations after baseline subtraction yield transition entropy changes, showing a decrease with increasing pressure (Supplementary Figure S8). This reduction is also observed for $x=100$\%, therefore indicating that this feature is intrinsic to SA and cannot be ascribed to a decrease in the molten SA molecules.

\begin{figure}[t!]
\centering
  \includegraphics[width=0.5\textwidth]{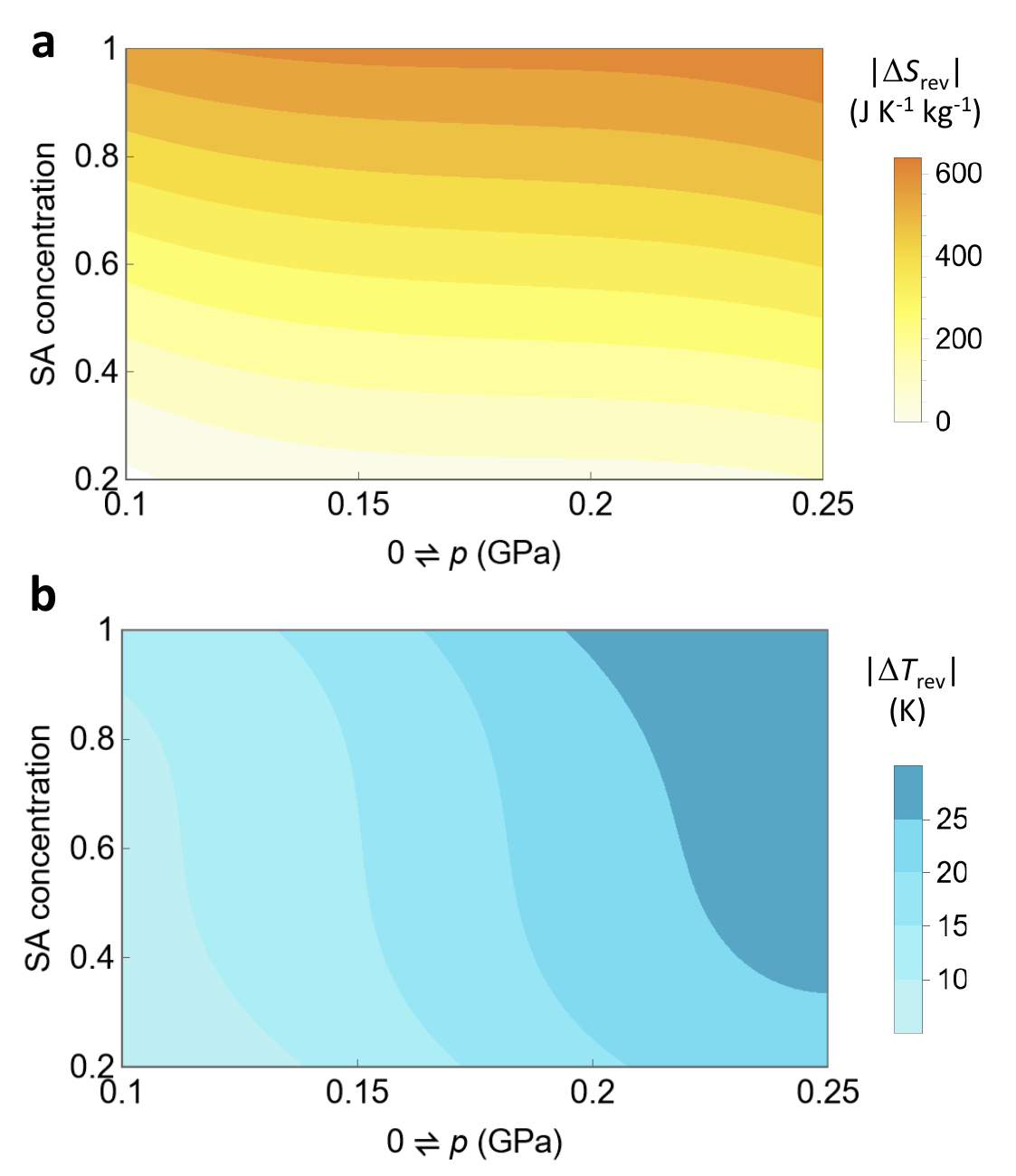}
  \caption{(a) Maximum reversible isothermal entropy changes and (b) maximum reversible adiabatic temperature changes displayed as a function of SA concentration and pressure changes.}
  \label{Fig5}
\end{figure}

\medskip
Next, isothermal entropy changes $\Delta S$ and adiabatic temperature changes $\Delta T$ driven by pressure changes (BC effects), per unit of mass of SA@MIL composite associated with the FOPT in SA, were determined using the quasi-direct method, whose calculation details can be found in Supplementary Section 3. Irreversible BC effects, i.e. those obtained under first compression/decompression processes, are shown in Supplementary Figure S11. Reversible BC effects, i.e. those that can be obtained upon repeated compression-decompression cycles, are shown in \textbf{Figure \ref{Fig4}} as a function of temperature for different pressure changes. For our composites, reversibility is achieved from pressure changes above $\sim$$50-75$~MPa. It is appropriate to highlight here the reversible BC response for $x=60$\% (Figure \ref{Fig4}c,h), close to the maximum concentration that could be adsorbed within the MIL porous structure, according to literature \cite{luan2016introduction}: under moderate pressure changes of $\simeq$100~MPa, colossal $|\Delta S_\text{rev}|=300$~J~K~kg$^{-1}$ and $\Delta T_\text{rev}\simeq$~6~K are reached. These values increase up to $|\Delta S_\text{rev}|=350$~J~K~kg$^{-1}$ and $\Delta T_\text{rev}\simeq$~20~K for large pressure changes of $\simeq$242~MPa. It is worth noting here that pressures of $\sim$100~MPa can be achieved using off-the-shelf compressors and are even smaller than those typically utilized in BC prototypes \cite{qian2024highly}. Notice also that multi-stage cascading designs \cite{torello2025finite} and active regeneration \cite{torello2020giant} can yield temperature spans much larger than $\Delta T_\text{rev}$ and therefore this parameter alone should not be a limiting factor.

\begin{figure}[t!]
\centering
  \includegraphics[width=0.48\textwidth]{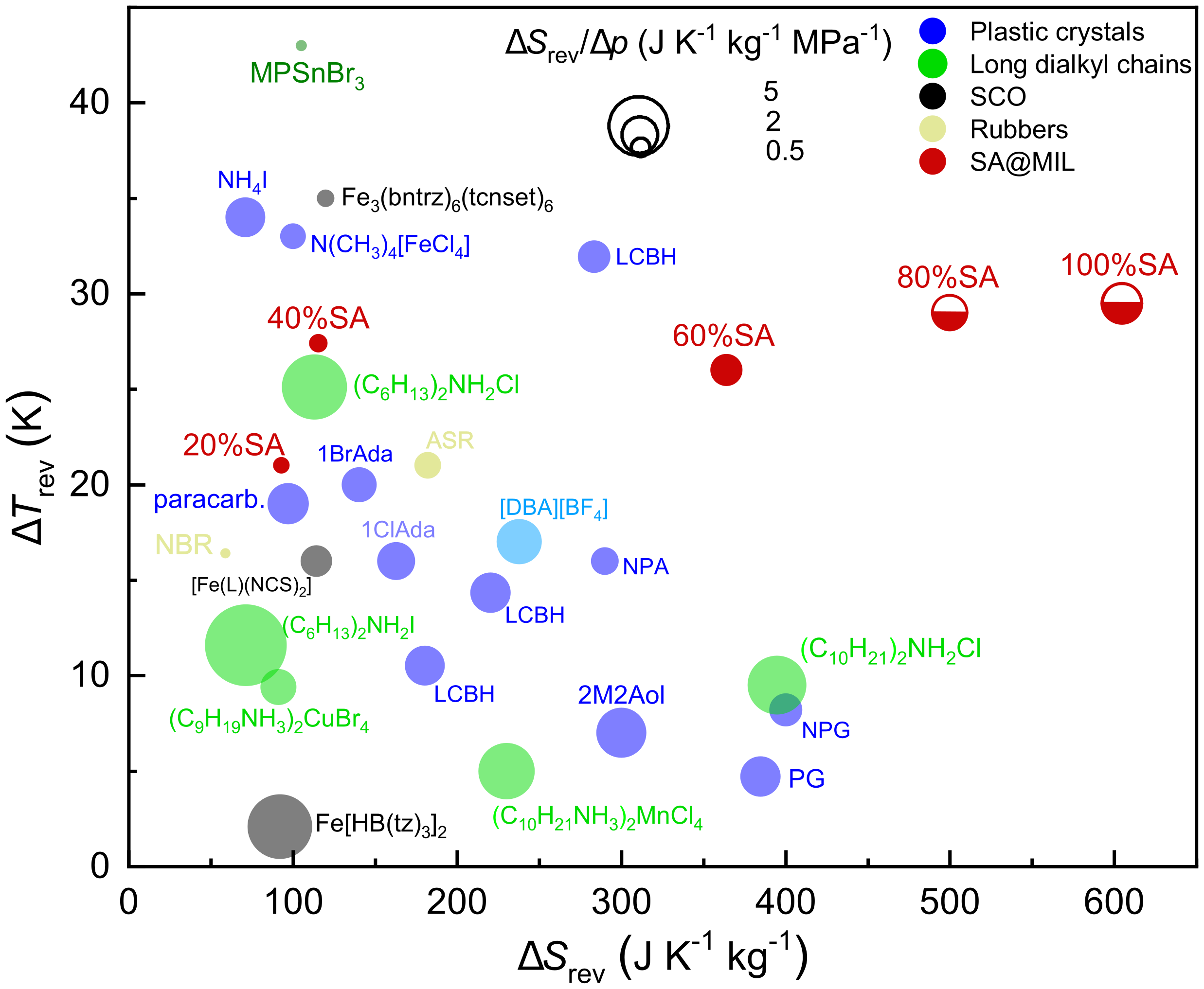}
  \caption{Comparison of reversible adiabatic temperature changes $|\Delta T_\text{rev}|$, isothermal entropy changes $|\Delta S_\text{rev}|$ and BC strength ($\Delta S_\text{rev}/\Delta p$, shown by the symbol size) among best performing BC materials when data are available. Colors refer to material families as indicated in the legend. Some symbols standing for the same material correspond to different applied pressures. Data taken from the following references: NPG [(CH$_3$)$_2$C(CH$_2$OH)$_2$], NPA [(CH$_3$)$_3$C(CH$_2$OH)] and PG [(CH$_3$)C(CH$_2$OH)$_3$] \cite{aznar2020reversible}; 1BrAda (C$_{10}$H$_{15}$Br) and 1ClAda (C$_{10}$H$_{15}$Cl) \cite{aznar2021reversible}; 2M2Aol (C$_{11}$H$_{18}$O) \cite{salvatori2022colossal}; (C$_6$H$_{13}$)$_2$NH$_2$Cl and (C$_6$H$_{13}$)$_2$NH$_2$I \cite{seo2024barocaloric}; 
  (C$_10$H$21$)$_2$NH$_2$Cl \cite{gao2024low};
  LCBH (LiCB$_{11}$H$_{12})$ \cite{zeng2024colossal}; p-carb (C$_2$B$_{10}$H$_{12}$) \cite{zhang2022colossal}; [DBA][BF$_4$] \cite{garcia2023structure}; CH$_3$PH$_3$SnBr$_3$ \cite{xu2025reversible}; [Fe(L)(NCS)$_2$] \cite{seredyuk2024reversible}; Fe$_3$(bntrz)$_6$(tcnset)$_6$ \cite{romanini2021giant}; Fe[HB(tz)$_3$]$_2$ \cite{seo2022driving}; NH$_4$I \cite{ren2022ultrasensitive};
 N(CH$_3$)$_4$[FeCl$_4$] \cite{salvatori2023large}.}
  \label{Fig6}
\end{figure}

\medskip
Maximum BC values as a function of SA concentration and applied pressure changes are displayed in \textbf{Figure \ref{Fig5}} as a colourplot derived from smoothed data in Figure \ref{Fig4}. Interfaces between value ranges make apparent that $\Delta S$ is strongly dependent on SA concentration (as it mostly depends on $\Delta S_\text{t}$) whereas $\Delta T$ is quite independent on SA concentration (as it mostly depends on $\frac{dT}{dp}$). We compare these BC effects of SA@MIL with other state-of-the-art BC materials in \textbf{Figure \ref{Fig6}}. The symbol size indicates the BC strength, defined as the ratio between $|\Delta S_\text{rev}|$ and the required pressure change. Therefore, big circles in the top right side of the panel are desirable. Our SA@MIL composites are displayed with pink symbols and half-filled symbols correspond to SA@MIL composites including a liquid contribution from molten SA that exceeds the loading capacity of MIL. The best performance within the solid state, achieved for $x=60$\%, compares very favourably with the state-of-the-art BC materials. Not only does this comparison demonstrate the potential of SL BC materials, but it clearly validates our strategy as a way to obtain solid systems with enhanced BC response.

\section{Conclusion}

In this work, we have shown the feasibility of confining melting transitions within metal-organic frameworks to harvest barocaloric effects for cooling and heating. This strategy provides the opportunity to take advantage of the superior heat storage capacity of molten phases without compromising the benefits of solid phases avoiding leakage and facilitating handling and recyclability. In particular, we have incorporated stearic acid inside MIL-101(Cr)-NH$_2$ in different concentrations. Using x-ray diffraction, scanning electron microscopy and high-pressure calorimetry, we have confirmed that colossal barocaloric effects associated with the melting of stearic acid are retained when encapsulated inside the MIL-101(Cr)-NH$_2$ pores, comparable to barocaloric effects per unit of composite mass at solid-solid phase transitions reported so far.
We have also found that the onset of BC effects is dependent on the concentration, with a reduction of up to 20 K between 20 and 100\% SA@MIL. In contrast to encapsulation within non-interacting microporous matrices, this behaviour emerges from the interplay between the stearic acid and the nanopore surface, providing opportunities to actively control BC properties by modifying parameters such as concentration and pore size, surface area and functionalisation. Our findings pave the way to explore multiple confinement combinations within the enormous pool of metal-organic frameworks and solid-liquid phase change materials with outstanding barocaloric performance in optimal temperature ranges.

\section{Experimental Section}

\subsection{Sample synthesis and preparation}
MIL-101(Cr)-NH$_2$, with ideal chemical formula [Cr$_3$O(OH)(H$_2$O)$_2$(C$_8$H$_5$O$_4$N)$_3$]$\cdot$nH$_2$O (n$\sim3$), was obtained by the synthetic method in Ref. \cite{Han2020}. For high-pressure differential thermal analysis, MIL-101(Cr)-NH$_2$ and quantities of SA were directly mixed and heated at 353~K for 30 minutes. Mixing was performed in different relative amounts of SA and MIL-101(Cr)-NH$_2$, yielding ratios of SA mass over composite SA@MIL mass $x=20$, 40, 60, 80 wt\%.

\subsection{Synchrotron diffraction}
Synchrotron powder diffraction was performed on the I11 beamline at Diamond Light Source, UK using $E_i = 25$~keV ($\lambda = 0.494556$~\AA) synchrotron radiation. The sample was packed in a 0.5~mm diameter borosilicate capillary. Temperature was controlled using an Oxford Cryostream and data was collected continuously with a ramp rate of 1 K min$^{-1}$ with the position sensitive detectors (PSD). Pawley refinements were performed using the GSAS-II software \cite{toby2013gsas}.

\subsection{Scanning Electron Microscopy (SEM)}
Images were taken using a Thermo Fisher Helios dual column plasma FIB-SEM. Samples were sputter coated with a thin layer ($\sim$few nm) of gold to eliminate sample surface charging.

\subsection{Porosimetry}
N$_2$ adsorption/desorption isotherms (77 K) were collected on a Micromeritics 3Flex instrument after activation at 423 K for 18 h under 10$^{-4}$ Torr vacuum. BET areas were calculated using the BETSI software package \cite{BET}.

\subsection{Calorimetry}
Differential scanning calorimetry (DSC) at atmospheric pressure was conducted using DSC Q100 (TA Instruments) by encapsulating a few mg of SA@MIL inside Al capsules and under scanning rates between 2 and 10 K min$^{-1}$. High-pressure differential thermal analysis (HP-DTA) was performed in a bespoke calorimeter operating within the range 0-0.3 GPa and 298 K - 473 K using  thermocouples arranged with Bridgman pistons as thermal sensors. A resistive heater was used to control heating runs at 3 K min$^{-1}$ whereas cooling runs were achieved in contact with an air stream. A few hundreds of mg of SA@MIL were encapsulated in Sn capsules with a drilled hole to insert a K-type thermocouple. The pressure-transmitting fluid was DW-Therm M90.200.02 (Huber).

\medskip

\medskip
\textbf{Acknowledgements} \par 
We thank Dr Catherine Walshe (University of Glasgow) for facilitating synthetic work and the China Scholarship Council (No. 202106300024) for funding. This work was supported by SGR-00343 Project (Catalonia), by Grant PID2023-146623NB-I00 funded by MICIU and by ERDF/EU, and is part of the Maria de Maeztu Units of Excellence Programme CEX2023-001300-M/ funded by MCIN/AEI/10.13039/501100011033. J.-L.T. is grateful to the ICREA Academia program. M.Z. is grateful for support from a Scottish University Physics Alliance (SUPA) Saltire Emerging Researcher Scheme. D.B. is grateful for support from a Leverhulme Trust Early Career Fellowship (No. ECF-2019–351) and a University of Glasgow Lord Kelvin Adam Smith Fellowship. For the purpose of open access, the author(s) has applied a Creative Commons Attribution (CC BY) licence to any Author Accepted Manuscript version arising from this submission.

\medskip
\textbf{Credit statement} \par
Conceptualisation: M.Z., D.B.; Investigation: M.Z., D.B., E.T.C., P.L., Y.W., F.R.B; Formal Analysis: M.Z., D.B., P.L.; Writing – original draft: M.Z., D.B., P.L.; Writing – review and editing: all authors; Supervision: D.B., P.L., R.S.F.; Funding acquisition: D.B., J.L.T., P.L.

Software, Data curation, Resources, Validation, Visualisation, Methodology, Project administration: Not relevant.


%
\bibliographystyle{MSP}
\bibliography{bib}

\end{document}